# Generative AI in K-12 Education: The *CyberScholar* Initiative

Vania Castro, University of Illinois Urbana-Champaign, USA
Ana Karina de Oliveira Nascimento, Universidade Federal de Sergipe, Brazil
Raigul Zheldibayeva, Zhetysu University named after I. Zhansugurov, Kazakhstan
Duane Searsmith, University of Illinois Urbana-Champaign, USA
Akash Saini, University of Illinois Urbana-Champaign, USA
Bill Cope, University of Illinois Urbana-Champaign, USA
Mary Kalantzis, University of Illinois Urbana-Champaign, USA

***Abstract:*** This paper focuses on the piloting of *CyberScholar*, a Generative AI (GenAI) assistant tool that aims to provide formative feedback on writing in K-12 contexts. Specifically, this study explores how students worked with *CyberScholar* in diverse subject areas, including English Language Arts, Social Studies, and Modern World History classes in Grades 7, 8, 10, and 11 in three schools in the Midwest and one in the Northwest of the United States. This paper focuses on *CyberScholar*'s potential to support K-12 students' writing in diverse subject areas requiring written assignments. Data were collected through implementation observations, surveys, and interviews by participating 121 students and 4 teachers. Thematic qualitative analysis revealed that the feedback tool was perceived as a valuable tool for supporting student writing through detailed feedback, enhanced interactivity, and alignment with rubric criteria. Students appreciated the tool's guidance in refining their writing. For the students, the assistant tool suggests restructuring feedback as a dynamic, dialogic process rather than a static evaluation, a shift that aligns with the cyber-social learning idea, self-regulation, and metacognition. For the teaching side, the findings indicate a shift in teachers' roles, from serving primarily as evaluators to guiding AI feedback processes that foster better student writing and critical thinking.

## Introduction

Artificial intelligence (AI) has emerged as a transformative force in society, impacting diverse domains of education, especially writing. Often considered an essential skill for college and employment readiness (National Commission on Writing, 2004), writing is crucial for engagement across disciplines, audiences, and is vital for participation in civic, professional, educational, and modern life activities (National Commission on Writing, 2003) (Cope & Kalantzis, 2000; Kalantzis et al., 2012 [2016]).

Despite its importance, national assessment programs in the US, such as NAEP, indicate a decline in writing performance among grade 8 students (National Center for Education Statistics, 2019). The present study aims to contribute to this critical gap by investigating the extent the Generative AI (GenAI) can assist middle and high school writing classes by providing formative feedback in diverse subject areas in Grades 7, 8, 10, and 11 in three schools in the Midwest and one in the Northwest of the United States.

This study is informed by formative assessment (Cope & Kalantzis, 2015; Kingston & Nash, 2011; Mahanan et al., 2021; Wiliam, 2011) and metacognition theory, that becomes more efficient when individuals monitor and reflect on their thought processes when receiving feedback (Cope & Kalantzis, 2015; Bielaczyc et al., 1995; Bransford et al., 2000; Kay et al., 2013; Lane, 2012; Magnifico et al., 2013; Winne & Baker, 2013).

This paper seeks to answer the following research question: To what extent does CyberScholar[1] have a potential to support K-12 students' writing in diverse subject areas? The assistant tool is a GenAI based writing assistant developed to support students' writing with formative feedback aligned with teachers' rubrics, instructional materials, and expectations. In doing so, the study aims to contribute to improving middle and high school students' AI literacy, especially considering that the tool's features provide students with the opportunity to receive AI feedback that directly connects to their teachers' work and expectations.

[1] CyberScholar is a Generative AI writing assistant tool developed at the University of Illinois Urbana Champaign as part of a U.S. Department of Education Small Business Innovation Research (SBIR) project in 2025.

## Literature Review

Prior studies have explored how GenAI can support student learning. Researchers such as Luckin et al. (2016) and Holmes et al. (2022) have noted their potential to enhance writing practice and provide individualized feedback on students' writing, supporting formative assessment. Similarly, AI tools, such as Grammarly and QuillBot, have shown promise in assisting students identify and correct grammatical mistakes more effectively (Godwin-Jones, 2022; Ghufron, 2019; Dembsey, 2017; Saeed, 2024). However, these studies are focused on higher education, leaving questions about how these tools could function in K–12 contexts.

Research on formative assessment, which provides real-time feedback during the learning process, has long been a goal in education (Cope & Kalantzis, 2015; Kingston & Nash, 2011; Mahanan et al., 2021; Wiliam, 2011). Nevertheless, its practical implementation has been inconsistent (Gurajala, 2020; Mahanan et al., 2021; Shepard, 2010). Recent studies point out that feedback generated by AI can help address this challenge by offering timely and personalized feedback to learners (Chan & Hu, 2023; Lee & Moore, 2024; Shi & Aryadoust, 2024). In this sense, Wan and Chen (2023) evaluated feedback provided by GPT-3.5 on middle school science writing. The study showed that students tended to rate the feedback by humans and GPT equally on correctness, but they all rated the feedback by GPT as more useful. These findings show the feasibility of using GenAI as an assistant to generate feedback for students.

A study by Steiss et al. (2023) investigated the use of GenAI for feedback in the K-12 setting. The study examined the use of ChatGPT for offering formative feedback on students' drafts. 200 students from grades 6-12 in a school in the USA participated in this research. The study involved a comparison of the feedback quality provided by both humans and AI, assessing them based on five measures: 1. adherence to criteria, 2. clarity in directions for improvement, 3. accuracy, 4. prioritization of essential features, and 5. employing a supportive tone. The results showed that differences between ChatGPT and humans were modest when considering the overall quality of feedback and time-savings, but human feedback was better than ChatGPT feedback for ⅘ elements of formative feedback.

Another recently published case study by Zheldibayeva et al. (2025) examined the impact of GenAI in writing assignments and the provision of formative feedback in K-12 context. The very first pilot study of CyberScholar, the tool mentioned in this paper, showed that it supported the development of students' writing in several ways.

While recent research demonstrates the potential of GenAI tools in writing instruction, other studies also note limitations and ethical concerns, especially in K–12 contexts, which include biases in AI-generated feedback, risks to student privacy, and inequities in access to the technology (Dong, 2023; Echols, 2023; Freiberger & Buchmann, 2024; Li et al., 2023; UNESCO, 2021, 2023).

The present paper introduces an additional dimension to the potential of GenAI for writing assistance by aligning AI-generated feedback with teachers' rubrics and classroom learning materials. This research analyzes the proposition of a new -- being tested -- GenAI tool, *CyberScholar,* that offers a potentially more consistent and pedagogically grounded method for formative assessment, directly connected to students' writing. The aim is to offer opportunities for collaborative and personalized feedback since the tool can be customized to fit teachers' requests and expectations based on the materials used and the rubric chosen for the evaluation of the specific content area and context. Students are able to present multiple versions of their work and ask the calibrated AI tool to provide feedback.

## Research Design

This study adopted a qualitative research design grounded in an interpretivist paradigm (Erickson, 1986), aiming to explore the extent to which the CyberScholar may support K–12 students' writing across diverse subject areas. Qualitative methods were chosen to capture the complexity of student interactions with GenAI, their experiences, and the contextual factors that shape writing development along with formative feedback.

An exploratory multiple case study approach (Mills, Durepos & Wiebe, 2010) was employed, focusing on four distinct school settings where the CyberScholar was being piloted. This design enabled a rich, contextualized understanding of how the platform functioned in varied educational environments and subject disciplines. Each case included a unique combination of grade levels, subject areas, and teacher practices, allowing for comparative insights while preserving the depth of individual cases.

The case study approach (Yin, 2014) was particularly suited to this research because it allowed for: an examination of real-world classroom use of GenAI in writing through the assistant tool, an exploration of students' perspectives on writing support and formative feedback, a special attention to the situated nature of writing practices across disciplines. By emphasizing context and meaning-making, the qualitative design supported the study's goal of uncovering how GenAI might influence writing processes, engagement, and outcomes in authentic educational settings.

The study also followed an agile methodology approach (Martin, 2009), incorporating iterative feedback loops to refine the implementation process, since there is also a component of tool development involved. Each development cycle in each school provided opportunities for participants to suggest adjustments and improvements, ensuring CyberScholar's adaptability to diverse classroom contexts. So this was part of the research pipeline that was a result of the research being conducted.

## Educational Context

The process of implementing CyberScholar in schools began by integrating teaching materials, writing prompts, and rubrics from the teachers through fine-tuning. This is done through Retrieval Augmented Generation (RAG) technology. i. e., materials are added to CyberScholar to supplement the information available in the underlying GenAI foundation model used in the platform. This ensures that feedback is aligned with educational standards and teacher expectations. Also, the materials were processed through AI prompt engineering to align feedback with classroom expectations. This is the main difference of the use of AI being evaluated by this research; it is used pedagogically, considering teachers' work, materials being used in class, and rubrics.

The initial trials were conducted in four public schools in the United States: three high schools and one middle school. The first, referred to as School A, is a public school located in an underserved area in the Midwest region of the country. The second, School B, is a well-resourced university laboratory high school, also in the Midwest. The third, School C, is a rural, mid-poverty middle school located in the Northwest. The fourth, School D, is a mid-poverty high school, also in the Midwest. These schools represent a diverse range of socioeconomic contexts: School A serves students from an underserved community with varied backgrounds, Schools C and D are situated in mid-poverty areas, and School B admits students from highly privileged social and economic backgrounds.

School A has 824 students enrolled in the 2024 school year, spanning grades 9 through 12. Its demographics show that approximately 35% of the student population is White, 30% is Hispanic, and 24% is Black. The highest enrollment is in Grade 9 (n=233) and Grade 10 (n=235), with a decline in Grade 11 (n=184) and Grade 12 (n=175). The teacher-student ratio is 13 to 1. The total dropout rate is 2.8%, with the highest incidence among Hispanic students. Regarding socioeconomic status, 99.6% of students qualify as low-income, meaning they are eligible for free or reduced-price lunches, live in substitute care, or belong to families receiving public aid (Illinois Report Card, 2023).

School B has 314 students enrolled in the 2024 school year, covering grades 8 through 12. According to school demographics, approximately 35% of students are Asian, 35% are White, 14% are Hispanic, and 7% are Black. The student distribution across grades is relatively even: Grade 8 (n=64), Grade 9 (n=64), Grade 10 (n=60), Grade 11 (n=65), and Grade 12 (n=61). The teacher-student ratio is 9 to 1. The school reports a zero dropout rate (Illinois Report Card, 2023). Additionally, data on low-income students, specifically, the percentage of students eligible for free or reduced-price lunches, living in substitute care, or from families receiving public aid—is redacted. As a result, this information is unavailable, unlike at School A (Illinois Report Card, 2023).

School C has a total enrollment of 69 students for the 2023-2024 academic year, spanning grades 6 through 8 (Alaska Department of Education and Early Development, 2024). The demographic breakdown reveals that 63.5% of students are considered part of a minority population, while 36.5% are White. Additionally, 58.7% identify as American Indian or Alaska Native, and 4.8% belong to two or more races. Data from 2025 reveals that the distribution of students across each grade level is relatively balanced: Grade 6 (n=23), Grade 7 (n=21), Grade 8 (n=19). The school maintains a teacher-student ratio of approximately 15 to 1 in 2025 (U.S News & World Report). There is no specific dropout rate available for this school in the Alaska Department of Education and Early Development records; only statewide data for Alaska's public schools is provided, though it is relatively low, 3.56% in 2023-2024. Regarding socioeconomic background, the school is classified as a mid-poverty school, with 37.5% of its students identified as low-income or economically disadvantaged.

School D has 1,825 students enrolled in the 2024 school year (Illinois Report Card, 2023). According to school demographics, approximately 45.5% of the student population is White, 26.8% is Hispanic, and 20.3% is Black. There is an even distribution of students across grade levels: Grade 9 (n=474), Grade 10 (n=468), Grade 11 (n=427), and Grade 12 (n=456,). The teacher-student ratio is 20 to 1. The dropout rate for the 2024 academic year is 2.5%. Additionally, 63.3% of students are classified as low-income, meaning they are eligible for free or reduced-price lunches, live in substitute care, or belong to families receiving public assistance.

## Participants

This paper is about research conducted in middle and high schools. It was designed considering the need to enhance the AI literacy of K-12 students. Focusing on that, researchers decided to test the extent to which CyberScholar has a potential to support K-12 students' writing in diverse subject areas. The investigation was conducted, particularly given that the capabilities of the tool, provide students with the opportunity to have AI feedback that directly connects to their teachers' work and expectations, thus enabling them to revise their writing, based on the AI review, initially guided and customized by their teachers. This ensures that students experience an AI review process based on reliable source materials connected to the interests and demands of the curriculum.

It was thus taken into account how AI can be harnessed to facilitate opportunities to improve equity in education by providing feedback and learning experiences to diverse populations in an inclusive and efficient manner. CyberScholar, as a scalable web platform, is designed to address resource disparities between school districts by meeting the personalized needs of students from diverse backgrounds across several factors: identity, cultural capital, and access to resources.

To have diverse populations involved, the flyer created to invite teachers to participate with their students was distributed across different states in the US. Despite this intention, the selection criteria were not purposive. Instead, convenience sampling was adopted, i.e., a non-probability sampling technique was used in which participants are selected based on their accessibility and proximity to the researcher, rather than through random selection. This was employed due to constraints related to time, resources, and logistics that limited the feasibility of the investigation. However, when teachers expressed their interest, the research team made sure to check whether diverse populations were included. While convenience sampling may be understood to compromise the representativeness of the sample and introduce potential bias, it remains a practical and widely accepted strategy in exploratory research, and qualitative inquiries, such as this one. By prioritizing ease of recruitment, researchers could efficiently gather preliminary insights, even if generalizability to the broader population is limited.

## Procedure

The study took place between August 2024 and January 2025, encompassing participant enrollment, trials, and data analysis. This timeline ensured seamless implementation, feedback collection, and iterative improvements. During this time, the recruitment of teachers and the software trials were conducted.

Following the research design, the first trials commenced after IRB approval. Recruitment letters, along with flyers, were sent to potential schools with assistance from the Director of School-University Research Relations at the University of Illinois Urbana-Champaign. The research team ensured compliance with the Student Online Personal Protection Act (SOPPA). Informational meetings were organized with teachers who showed interest to explain the study, discuss the proposed writing topics aligned with the curriculum, and outline the purpose of the CyberScholar tool, especially the component. These meetings also emphasized the potential benefits of the tool in enhancing students' writing skills and supporting teachers' workload. To ensure proper implementation, training sessions were held for both teachers and students, focusing on how to effectively use the tool.

Upon receiving students' and parents' permission, accounts on CyberScholar were created for students, using their school e-mails provided by the teachers. Only then the implementation phase took place.

## Data Collection

After receiving participants' permission, pre-surveys were sent to both teachers and participating students. The next step was teacher's training in order to provide familiarization with the tool. On the following day scheduled with the teacher, students' training was provided, having the research team and the teacher's guidance. Later, the implementation itself took place. During this phase, observations were made in classrooms to evaluate how the teachers integrated and utilized the CyberScholar, while participating students provided feedback on its effectiveness in improving their writing assignments. For students who chose not to participate in the study, alternative assignments were provided by the teachers. These assignments, which did not require the use of the tool, ensured that all students engaged in meaningful learning activities aligned with the curriculum without any disadvantage to those who opted out.

Throughout the research process, the team conducted regular check-ins with the participating teachers to provide ongoing support. At the conclusion of the implementation phase, an online post-survey was distributed to participating students and teachers to collect feedback. Additionally, interviews were conducted with both students and teachers to gather more details.

## Data Analysis

Data was analyzed based on the following research question: To what extent does CyberScholar have a potential to support K-12 students' writing in diverse subject areas? In order to observe the findings, codes were initially attributed to the data. For the analysis, the understanding was that the code would refer to a term or concise phrase developed by the team of researchers to represent the core idea, or significant theme within a segment of linguistic data. Coding would then assign specific interpretive meaning to individual units of data, facilitating the identification of patterns, the organization of content into categories, and the advancement of analytical interpretation (Saldaña, 2013).

Although 121 K-12 students participated in all research steps, generating AI feedback and making revisions to their texts, a total of 54 students and 03 teachers from the four schools completed the post-survey, 8 students and 1 teacher participated in the interviews. In order to analyze the data, following Saldaña (2013), initial coding was developed. It involved segmenting qualitative data into distinct units for detailed examination, enabling the identification of patterns through comparative analysis of similarities and differences. This methodological approach is broadly applicable across the spectrum of qualitative research designs, and that was chosen for this particular study.

After this initial stage, the second cycle coding involved pattern coding, which is about the recognition of data segments that share similar initial codes and the systematic organization of these segments into thematic groupings. This technique was selected for its suitability in second cycle coding, particularly in facilitating the emergence of overarching themes and supporting the interpretive pursuit of underlying explanations within the dataset (Saldaña, 2013).

## Findings

Of the respondents to the post-survey, 36 students explicitly stated that they would like to use the CyberScholar tool again for writing assignments. A total of 11 students across the four schools indicated in the post-survey that they were hesitant about using the tool again. Initially knowing most students would approve of the GenAI tool to support their writing assignments, the next step was to work on first cycle coding and then second cycle coding. Based on the researchers' analysis and following pattern coding, some themes emerged, which are presented below.

According to the data analyzed, AI assistant tool has the potential to support K-12 students' writing through: 1. The Provision of Detailed and Specific feedback; 2. The Interactive Nature of the GenAI Tool; 3. The Development of Students' Perception of Improvements in Writing Quality; 4. The Delivery of Ratings Connected to Rubric Criteria.

### 1. *The Provision of Detailed and Specific Feedback*

This theme highlights aspects that students praised, particularly that the CyberScholar provided detailed and specific feedback on the writing assignments. In the post-survey, 19 students across the four schools explicitly mentioned in their comments positive opinions about the detailed and specific feedback. Students noted that the tool's feedback allowed them to identify specific areas for revision, which contributed to their overall writing development. Some students highlighted that the tool provided concrete ways to fix issues, helping them understand their mistakes and how to improve their writing, especially in areas in need for more nuanced and sophisticated expression.

One student highlighted that the AI review corrected her text in chronological order, which made it easy for her to make corrections. Another student appreciated the tool's organization and the use of basic bullet points. A third student noted that the AI feedback was focused and made it easy to locate specific information or areas needing improvement, particularly in the revision process. The following quotes are examples of these views:

> *"[CyberScholar] Told me exactly what I could do to improve. Got me to write it more as a college-like paper."* (Student 1 - School A)
>
> *"I liked that it gave us a very detailed description on how our arguments could be presented clearly."* (Student 15 - School B)
>
> *"I liked how it told me exactly what I needed to fix by putting the suggestion in quotes."* (Student 17 - School C)
>
> *"I liked that it offered recommendations on how to change sentences as opposed to simply saying to change it."* (Student 44 - School D)
>
> *"It did not just explain to me what I needed to do, but it also gave me specific examples of how to make my writing better."* (Student 54 - School D)

Some students decided to run the AI review again after their revisions. They noticed improvements in their scores, which meant that the assistant tool had provided them with valuable formative feedback to enhance their writing development. For example, Student 2 from School A improved these following items from the rubric "Compare" and "Contrast" score from 0 to 2. After the first round of CyberScholar feedback, they were able to identify some similarities and differences between the first and the last revised texts, which resulted in a higher rating and improved writing. It was found that the feedback from the AI tool was useful for enhancing their writing.

### *2. The Interactive Nature of the GenAI tool*

This theme presents students' appreciation for the interactive feature of the CyberScholar tool, which allowed them to engage in a dynamic conversation with the tool during the revision process. Six students across the schools explicitly praised the interactivity of the GenAI tool. It is important to note that this functionality was added as the research progressed and was available only for Schools C and D. Students asked follow-up questions for further clarification, making the formative feedback process more engaging and useful. This functionality enabled students to be more active in the learning process rather than passively receiving comments. This can be seen in the following quotes:

> *"I liked how you could talk back to the AI. There's software out there that give you feedback, but this system is unique because you can have a sort of conversation with the AI, allowing for a ton of diversity."* (Student 30 - School D)
>
> *"I liked most how it gave good grammar helps and was able to present its data in a digestible way. It was also pretty apt to my questions and gave good answers."* (Student 32 - School D)
>
> *"I like how it responds to my questions. Especially when I asked "Can you further elaborate what I need to fix in this section?" It gave me good feedback and explained thoroughly when I asked it to elaborate further"* (Student 40 - School D)

### *3. The Development of Students' Perceptions of Improvements in Writing Quality*

This theme highlights the student key benefits of using the CyberScholar tool and how it helped in their writing. 23 students across the four schools explicitly mentioned that the tool helped them realize they could improve their writing. They mentioned how it provided specific recommendations for their texts, such as refining grammar, conclusions, enhancing transitions, and making better connections between points. Some students noted that the tool helped make their writing sound more professional by suggesting appropriate word choices and provided recommendations for sentence structure, and clarity. The following quotes demonstrate what students noticed on their writing improvements:

> *"We learned that our conclusions should not just simply restate what we were talking but, especially in a History assignment. We should also try to explain the significance of this project."* (Student 15 - School B)

> *"I wrote down better transitions."* (Student 18 - School C)

> *"It gave me more filler words to make it sound more professional."* (Student 20 - School C)

> *"I liked that it would give precise examples of what was wrong, and present ways to fix it."* (Student 30 - School C)

> *"It helped me clear up some sentences that sounded a bit too clunky and helped me re write them."* (Student 36 - School D)

> *"Helped refine my grammar."* (Student 27 - School D)

### *4. The Delivery of Ratings Connected to Rubric Criteria*

This theme highlights participants' satisfaction with the ratings provided by the CyberScholar tool and how the tool applied rubric criteria to their texts even though it revealed divided opinions among the participating students as to how to visualize the ratings.

Some students pointed out that the tool provided them with structured and consistent feedback by assigning ratings and scores based on evaluation criteria created by their teachers. The tool's alignment with teachers' rubric was seen as positive and understood as a big difference compared to their individual uses of GenAI in real life. Therefore, students appreciated that the AI explicitly referenced rubric categories like "writing mechanics." This suggests the tool made the assessment criteria more visible and understandable, which is often a challenge in traditional feedback. Besides that, the way the tool was programmed, automated feedback mimicked teacher expectations, bridging the gap between abstract rubric language and concrete writing performance. This reflects a shift from generic feedback, often common in uses of GenAI tools to precision-guided revision, which is a hallmark of effective formative assessment, the aim of the tool being tested.

However, the way the ratings were displayed -- through stars -- generated different opinions. For example, one student explained that the ratings helped him understand how much to improve transitions and accurately reflected the quality of his writing, making the stars a true representation of his journal entries. Similarly, another student appreciated the stars as a rating system because they motivated her to earn more stars in the next round of AI feedback than she had previously received. During the interviews with students from School B, they mentioned how they appreciated the star-based grading system for providing a clear visual understanding of their results. One of the students mentioned that this encouraged her to work even harder. The following quotes are examples of these views:

*"I liked the star ratings and how in-depth the explanations were"* (Student 2 - School A)

*"I liked the fact that the AI went over specific sections of the rubric, like the writing mechanics, and gave a rating out of four depending on how well the essay was written."* (Student 37 - School D)

*"Yes, I liked how it rated my essay in each category so I could see where I needed to focus my attention on which was nice. This way I could see the improvement of my writing each time I submitted a revised version."* (Student 26 - School D)

On the other hand, some participants found the ratings inconsistent, noting that they sometimes received different scores for the same submission without making any changes.

*"The ratings were kinda random, like I told the AI to reevaluate and it gave me a better score even though I didn't make any changes."* (Student 27 - School D)

*"I did not like that when I put in the same paper twice it gave me two very different rankings. It made me unsure of what score my paper should have actually gotten and made things more difficult."* (Student 35 - School D)

### *5. Other less common themes*

Other less common themes emerged. According to the data, CyberScholar also has the potential to support K-12 students' writing through: the understanding of it as a resource for self-directed learning, the provision of personalized and quick feedback, the possibility of saving students' time, as the tool quickly identified areas for improvement in students' texts, and provided fast feedback.

Student 32 from School D explained why he would use the tool again:

*"[...] it really helped with the flow of my writing, it also gave me pretty good ideas on where I was good and then where I was lacking in relation to the rubric, thankfully it was also an independent and unbiased source so I would be able to trust it. It is also easier and quicker to work with than a peer or use self-analysis."*

## Discussion

This study investigated the extent to which the GenAI assistant tool, CyberScholar, can support K–12 students' writing in diverse subject areas. The findings indicate that participating students perceived the tool as a valuable help in their writing development, especially in receiving detailed and specific feedback connected to their teachers' rubric, engaging in interactive clarification, and developing their writing quality.

In correlation with recent studies on the practical implementation of GenAI Tools in writing (Godwin-Jones, 2022; Ghufron, 2019; Dembsey, 2017; Saeed, 2024), participants in this study also observed grammar improvements and corrections in their texts. However, unlike other tools being studied, CyberScholar incorporated teaching resources, such as books used by the teacher and PowerPoint slides, to supplement the AI feedback provided to students. This made the feedback more closely aligned with educational standards and teacher expectations across different subject areas. Also, assessment criteria were more transparent and familiar to students.

The provision of detailed, specific, and timely feedback in the students' texts, reflects the key aspects of formative assessment discussed by Cope & Kalantzis (2015), Chan & Hu (2023), and, Lee & Moore (2024). Similar to findings by Wan and Chen (2023) and Steiss et al. (2023) the participants also evaluated the quality of the AI feedback, where they point out as actionable, especially when feedback pinpointed specific areas for revision and offered examples of improved phrasing. Some students re-submitted revised drafts to observe improvements in their rubric scores, tracked their progress demonstrating iterative engagement that aligns with the idea of the AI and human feedback learning model by Cope & Kalantzis (2019, 2023). This model highlights the combination of formative assessment, cycles of cognition, and metacognition (reflecting on the writing process).

Students also appreciated the possibility to ask follow-up questions in CyberScholar, a feature that distinguished the tool from other static feedback tools. This functionally potentially increases student agency and creativity during the revision process.

The interactive functionality of the AI tool with humans during the provision of a feedback aligned to what participants were studying, leads to "cyber-social relations of meaning" and, more specifically, "cyber-social learning" for educational contexts (Cope, Kalantzis, and Zapata, 2025). This means that learning happens through dynamic interactions between human agency and AI-generated feedback, and meaning is co-constructed rather than passively received (Cope & Kalantzis, 2023). This connection also informed the team's choice of the name CyberScholar for the tool.

The findings of this study suggest that when designed with clear instructional alignment and opportunities for student interaction, GenAI tools can play a meaningful role in formative assessment. However, to fully realize this potential, it is important to consider ethical implications. In collaboration with humans, AI can help address critical issues in education, including unequal access, the need for personalized learning, and teacher and student well-being.

## Final Considerations

The main objective of the present paper was to investigate the extent to which the GenAI tool -- CyberScholar -- can support K–12 students' writing in diverse subject areas. The results show that the tool has substantial potential to improve students' writing and support teachers in their instruction. It is important to note that participants acknowledged the potential to support writing but not to replace teacher feedback.

The AI assistant allowed timely and detailed feedback integrated into the teachers' specific rubric, content, and subject. For students, the CyberScholar tool suggests restructuring feedback as a dynamic, dialogic process rather than a static evaluation, a shift that aligns with the cyber-social learning idea, self-regulation, and metacognition.

On the teaching side, the findings demonstrate a potential evolution in teachers' roles, moving from sole evaluators to managers of AI feedback loops, in ways that deepen students' writing and critical thinking. The study also suggests a need for classroom designs that support iterative revisions, inquiry-based dialogues, and personalized human–machine learning pathways.

CyberScholar is under research and development and will continue to incorporate improvements suggested by participants along with observations from the research team for the next phases of implementation. Iterative cycles of development and evaluation will continue, as they are essential to realizing its full potential as a transformative educational tool in real-world educational settings.

## Challenges and Limitations

The present study has provided some significant findings to support student writing, challenges and limitations should be acknowledged.

The piloting of CyberScholar was carried out with a small sample of schools and classrooms (n=121 students, n=4 teachers, and limited post-survey response rates (n=54), which may limit generalizations of findings to represent all K–12 learning environments. The results are exploratory, and that more research is necessary to confirm broader applicability.

Future studies using the AI feedback tool should expand the sample size and include a larger number of schools from different geographic regions, socioeconomic contexts, and subject areas.

The study emphasized the role of CyberScholar in providing AI feedback on students' texts as a complement to teacher feedback. Further investigation is needed to ensure that AI use does not reduce opportunities for critical thinking and that ethical considerations are fully addressed.

## AI Acknowledgment

The authors acknowledge the use of Grammarly (paid version) and ChatGPT (GPT-5 free version) for grammar correction and language refinement. Grammarly was utilized to enhance clarity, coherence, and correctness. The prompts used include “Correct grammar" and “Proofread for academic writing but don't change the original meaning or structure of the text”. While the authors acknowledge the usage of AI, they maintain that they are the sole authors of this article and take full responsibility for the content therein, as outlined in COPE recommendations.

### Informed Consent

The authors have obtained informed consent from all participants.

### Conflict of Interest

The authors declare that there is no conflict of interest.

## ABOUT THE AUTHORS

**Vania Carvalho de Castro**: Teaching Assistant Professor in University of Illinois Urbana-Champaign, College of Education, Department of Education Policy, Organization and Leadership, Champaign, Illinois, United States
Email: vaniac@illinois.edu

**Ana Karina de Oliveira Nascimento**: Professor, Universidade Federal de Sergipe, Brazil, University of Illinois Urbana-Champaign
Postdoctoral Fellowship by the National Council for Scientific and Technological Development – CNPq, Brazil
Email: akcoliveira@academico.ufs.br

**Raigul Zheldibayeva**: Researcher, Zhetysu University named after I. Zhansugurov, Kazakhstan, University of Illinois Urbana-Champaign, Bolashak International Scholarship, Kazakhstan

Email: raigulz@illinois.edu

**Duane Searsmith**: Duane Searsmith, AI Developer and Researcher, University of Illinois Urbana-Champaign, College of Education, Department of Education Policy, Organization and Leadership, Champaign, Illinois, United States
Email: dsearsmi@illinois.edu

**Akash Saini**: PhD student, University of Illinois Urbana-Champaign, College of Education, Department of Education Policy, Organization and Leadership, Champaign, Illinois, United States
Email: aksaini2@illinois.edu

**Bill Cope**: Professor, University of Illinois Urbana-Champaign, College of Education, Champaign, Department of Education Policy, Organization and Leadership, Illinois, United States
Email: billcope@illinois.edu

**Mary Kalantzis**: Professor, University of Illinois Urbana-Champaign, College of Education, Department of Education Policy, Organization and Leadership, Champaign, Illinois, United States
Email: kalantzi@illinois.edu

This section only supports the fields listed above. A corresponding author must be indicated for works with two or more authors. Do not provide additional information, such as your ORCID iD, or a full biography in this section. Instead, please update your CGScholar profile: https://cgscholar.com/cg_support/en/docs/100.